\documentclass[12pt]{article}
\usepackage{graphicx}

\usepackage[usenames]{color}
\usepackage{makeidx}

\begin{document}

\title{{\bf Entanglement in parametric interactions in presence of phase-mismatch }}

\author{Ana M. Martins \\
%EndAName
Departamento de F\'{i}sica, Instituto Superior T\'{e}cnico, 1049-001
Lisboa, Portugal }

\date{\today}

\maketitle

\begin{abstract}
We investigate the influence of the phase-mismatch in the emission rate of photon-pairs and in their degree of entanglement in nondegenerate parametric processes. The Hamiltonean describing this processes contains explicitly the phase-mismatch and is solved exactly. The time dependent bosonic operators obey to {\it generalized Boguolibov transformation}. We show that the average number of photons produced in the interaction, the degree of squeezing as well as the degree of entanglement decrease with increasing phase-mismatch.
\end{abstract}

PACS number(s) 42.65.Lm, 03.67.a, 42.50-p.

%\widetext

\newpage

\section{Introduction}

 A significant number of Quantum Computation and Quantum Information protocols use the quadrature amplitudes of the quantized electromagnetic field \cite{Braunstein1998}, \cite{Plenio2003}. The essence of all these protocols rely in the possibility of producing entangled photons in optical nonlinear mediums  \cite{Ou1992}.

In particular, the nondegenerate parametric amplifier (NPA) with injected signal, and the spontaneous parametric downconversion (SPD) generate squeezed states of light \cite{Reid1988,Kimble1986,Giacobino1987} and are paradigmatic processes in the realm of the continuous variable (CV) entanglement \cite{Braunstein2005}. 

The efficiency of any nonlinear optical process depends strongly on this wave-vector mis- match. The emission is brightest if the various fields are coherent over the full length of the
8

Dispersion in the nonlinear material between pump, signal, and idler waves hereby plays an inhibiting role and should be eliminated. This can be achieved with the use of birefringent nonlinear crystals. In such crystals perfect phase matching can be attained by appropriately orienting the crystal axes, and the wave vectors and polarization vectors of the input fields. However, it is often the case that the strongest nonlinearities cannot be addressed in this man- ner. Consequently, considerable effort has been devoted to develop materials that combine perfect phase matching with a strong nonlinear response, resulting in a multitude of quasi- phase-matched and periodically-poled materials [88, 89].

It is then worth  to inquire how the quality of entanglement achieved in a parametric amplifier and in a spontaneous parametric downconversion can be affected when the phase matching condition cannot be completely achieved in actual experiments. It is known from classical nonlinear optics \cite{Bloenbergen1998} that the existence of a phase mismatch (i.e. a nonzero $\Delta  {\vec k}$), that occurs during the propagation in the nonlinear medium, modify the power of the signal and of the idler. Therefore, absence of perfect phase matching will certainly affect  efficient amplification, degree of squeezing and entanglement of the interacting modes.

In this paper we generalize the usual analysis of the nondegenerate parametric processes \cite{Reid1989,LOUISELL1961} to take into account the influence of the phase mismatch in the degree of entanglement. In the classical version of this problem \cite{Bloenbergen1998} the phase-mismatch is mathematically equivalent to a detuning in the frequency of the interacting modes.  The corresponding quantized version is described by an extension of the usual Hamiltonean \cite{LOUISELL1961,Reid1989} which is no more time independent in the interaction picture. We compute the exact solution of the Heisenberg equations of motion to find the time dependent bosonic operators that explicitly depend on the phase-mismatch.

This model also describes nondegenerate parametric interactions of waves with finite spectral bandwidth \cite{Martins1988} ,this is, the energy conservation rule is obeyed within the imprecision of the finite bandwidth of the pump field. 

The characterization of the entanglement will be done using the well stablished techniques  of continuous variables Gaussian states developed for bipartite systems \cite{Illuminati04}, this is, we use the formalism of the covariance matrix  to compute the Von Neumann entropy of each mode as well as the logarithmic negativity of the composite system, for different initial quantum states. In all cases we prove that the degree of entanglement decreases with increasing mismatch or with finite bandwidth.

The plan of the paper is the following: In Sec.II we introduce the Hamiltonean of the system with phase-mismatch, solve the Heisenberg equations of motion for the bosonic operators and compute the average number of generated photons. In Sec. III we compute the covariance matrix of the system for different initial states and determine in each case the degree of entanglement of the system. Finally in Sec.IV we present some conclusions.

\section{System dynamics }

In the NPA a pump mode with frequency $ 2 \omega $ and wavevector $2 {\vec k}$  interacts in a nonlinear medium with a mode of frequency $\omega_1$ and  wavevector ${\vec k}_1$ (the signal ) which grows. In the process an auxiliary mode of frequency $\omega_2$ and  wavevector ${\vec k}_2$ ( the idler ) is created. The SPDC differs from the NPA because the only input to the nonlinear medium is the pump mode and downconversion to the lower-frequency modes $\omega_1$ and $\omega_2$ is spontaneous. We consider that the incident pump mode is intense and can be treated classically as a field of complex amplitude ${\cal A} = |{\cal A}| e^{-2 i \omega t}$, only modes $1$ and $2$ are described by the bosonic operators $\hat{a}_1$ and $\hat{a}_2$. 

In both cases the three modes obey the energy conservation rule $2 \omega = \omega_1 + \omega_2$ and the phase mismatch $ \Delta { \vec k}= 2 {\vec k} -  {\vec k_1}-{\vec k_2} \neq 0$.  The Hamiltonean describing both processes and that generalizes  \cite{LOUISELL1961,Reid1989}, is given by \cite{Bloenbergen1998}
\begin{equation}\label{hamiltonean}
{\hat H} = \hbar \omega_1 \hat{a}_1^\dag \hat{a}_1 +     \hbar \omega_2\hat{a}_2^\dag \hat{a}_2 -    \hbar  g (e^{-2 i (\omega   + \delta) t}\hat{a}_1^\dag \hat{a}_2^\dag+ e^{2 i (\omega   + \delta) t}\hat{a}_1 \hat{a}_2)
\end{equation}
where the interaction time $t$ is the propagation time in the nonlinear medium therefore proportional to the travelled distance $z$. The phase factor $e^{2 i \delta t}$ with $ \delta > 0$ results from the mismatch along $z$ and the coupling constant $g$ is proportional to the second order susceptibility of the medium and to the amplitude of the pump. 

The Hamiltonean (\ref{hamiltonean}) can also describe the interaction between a classical wave of central frequency $2 \omega $ and bandwidth $2 \delta$, with two quantized modes of frequencies $\omega_1$ and $\omega_2$, and such that the energy conservation rule $2 \omega = \omega_1 + \omega_2$ is obeyed. 

The Heisenberg equations of motion for modes $1$ and $2$ are
\begin{equation}\label{Heis}
\frac{d}{dt} 
\left( \begin{array}{cc}
\hat{a}_1\\
\hat{a}_2^{\dag}\\
\end{array}
\right)
= \left(
\begin{array}{cc}
-i \omega_1  & ig e^{-2 i (\omega   + \delta) t}\\
-ig e^{2 i (\omega   + \delta) t}& i \omega_2 \\
\end{array} 
\right)
\left(
\begin{array}{c}
\hat{a}_1\\
\hat{a}_2^{\dag}\\
\end{array} 
\right)
\end{equation}
jointly with their hermitian conjugates, which are decoupled from these two. When the mismatch is zero the integration of these equations is done in the interaction picture. However, this usual approach doesn't work in the case of a nonzero mismatch because it doesn't remove the time factor $e^{-2 i  \delta t}$ from the Heisenberg equations. In order to obtain an autonomous system of equations we are going to define new annhilation operators $\hat{b}_1$ and $\hat{b}_2$, by the following canonical transformation
\begin{equation}\label{FA}
\left(\begin{array}{c}
\hat{b}_1\\
\hat{b}_2^{\dag}\\
\end{array}
\right)=
\left(\begin{array}{cc}
e^{ i (\omega   + \delta) t}& 0\\
0&e^{- i (\omega   + \delta) t}\\
\end{array}
\right)
\left(\begin{array}{c}
\hat{a}_1\\
\hat{a}_2^{\dag}\\
\end{array}
\right)
\end{equation}
Physically this corresponds to move to a reference frame that rotates with the frequency $(\omega   + \delta)$.
The Heisenberg equations for the new  operators $\hat{b}_1$ and $\hat{b}^{\dag }_2$ are then given by
\begin{equation}\label{Heis1}
\frac{d}{dt}
\left(
\begin{array}{c}
\hat{b}_1\\
\hat{b}_2^{\dag}\\
\end{array}
 \right)
= \left(
\begin{array}{cc}
i (\omega - \omega_1+ \delta)& ig \\
-ig & i (\omega - \omega_1- \delta)\\
\end{array}
\right)
\left(\begin{array}{cc}
\hat{b}_1\\
\hat{b}_2^{\dag}\\
\end{array} 
\right)
\end{equation}
They are linear time independent differential equations and can be readily integrated. The only solution of eq.(\ref{Heis1}) with physical interest corresponds to the condition $ \delta < g $ where the mismatch is small compared with the coupling parameter. By inverting eq.(\ref{FA}) and using the solution to eq.(\ref{Heis1}) we obtain finally the time dependent bosonic operators
 \begin{equation}\label{time1}
\hat{a}_1(\tau)= e^{-i \tau \omega_1^{\prime}(y) } \left[ \left( C(x\tau)+  \frac{i y}{(1-y^2)^{1/2}} S(x \tau) \right) \hat{a}_{10} + \frac{i}{(1-y^2)^{1/2}} S(x \tau)  \hat{a}_{20}^{\dag} \right]
\end{equation}
\begin{equation}\label{time2}
\hat{a}_2^{\dag}(\tau)= e^{i \tau \omega_2^{\prime} ( y)} \left[ - \frac{i}{(1-y^2)^{1/2}} S(x \tau )  \hat{a}_{10} +  \left( C(x \tau) -  \frac{i y}{(1-y^2)^{1/2}} S(x \tau) \right) \hat{a}_{20}^{\dag} \right]
\end{equation}
where $\tau = gt$ is an effective dimensionless interaction time, $y= \delta /g$ is the dimensionless mismatch parameter with values in the interval $0 \leq  y < 1$, the effective frequency of oscillation of mode $j$ is $\omega_j^{\prime} (y)=  \omega_j / g + y$. 

The system of equations (\ref{time1}) and  (\ref{time2}) is a kind of {\it generalized Boguolibov transformation} where $C(x t)= \cosh ( x gt) $, $  S(x t)= \sinh ( x gt) $ are the hyperbolic functions, $x=  (1-y^2)^{1/2}$ and $\hat{a}_{10}= \hat{a}_1(0)$ and $\hat{a}_{20}^{\dag}=\hat{a}_2^{\dag}(0)$ are the bosonic operators at time $t=0$.  

Equations (\ref{time1}) and (\ref{time2}) extend the well known result \cite{LOUISELL1961} to take into account the influence of the phase-mismatch. The corresponding bosonic operators derived in \cite{LOUISELL1961} are obtained by making $y =0 $ in the last equations. 

The argument of the hyperbolic functions is the squeezing parameter,  
\begin{equation}\label{squeezing}
r(y) =gt  (1-y^2)^{1/2},
\end{equation}
which is a generalization of the usual squeezing parameter $r= gt $. It depends not only on the interaction time $t$, but also on the  mismatch. For a given interaction time the intensity of the squeezing parameter decreases with increasing mismatch.

From the Heisenberg equations  (\ref{Heis}) we obtain the following integral of motion
\begin{equation}\label{Inv}
\langle {\hat n}_1  (t)  \rangle  - \langle {\hat n}_2 (t)  \rangle =  { \bar n}_{10}  -  { \bar n}_{20}
\end{equation}
which is a generalization, for nonzero mismatch, of the analog integral of motion derived in \cite{LOUISELL1961}.  We have introduced in the previous equation the initial mean number of photons ${ \bar n}_{10}  $ and $ { \bar n}_{20}$. 
This integral of motion says that each time the pump creates a photon in mode (1) it also creates a photon in mode (2) therefore, the difference of photons between these modes at any time $t$ is equal to their difference at the initial time $t=0$ and that it is independent of the phase-mismatch. For uncorrelated input vacuum states the average number of photons in each mode is initially, $ { \bar n}_{10} = { \bar n}_{20} =0$, jointly with equality (\ref{Inv}) it determines that the average number of photons  in the signal is always equal the the mean number of photons in the idler.

From the time dependent bosonic operators given in eqs.(\ref{time1}) and (\ref{time2}) we compute the average photon number in mode $j (=1,2)$,
\begin{equation}\label{n1}
\langle {\hat n}_j  (\tau)  \rangle =  \frac{1}{1-y^2} [ C^2( x \tau) -y^2) {\bar n}_{j0} + S^2(x \tau)(  { \bar n}_{j+1,0} +1) ]
\end{equation}
Because the pump amplitude $\cal A$, is treated as constant, the solution to the parametric interaction cease to be valid when the  number of pair of photons is such that determines an appreciable depletion of the pump mode, this limits our calculation to the case $\langle {\hat n}_j  (\tau)  \rangle << | { \cal A }|^2$ and $\tau = g t <1$.

\begin{figure}[htb]
\centering
\includegraphics[height=50mm,keepaspectratio=true,angle=0]{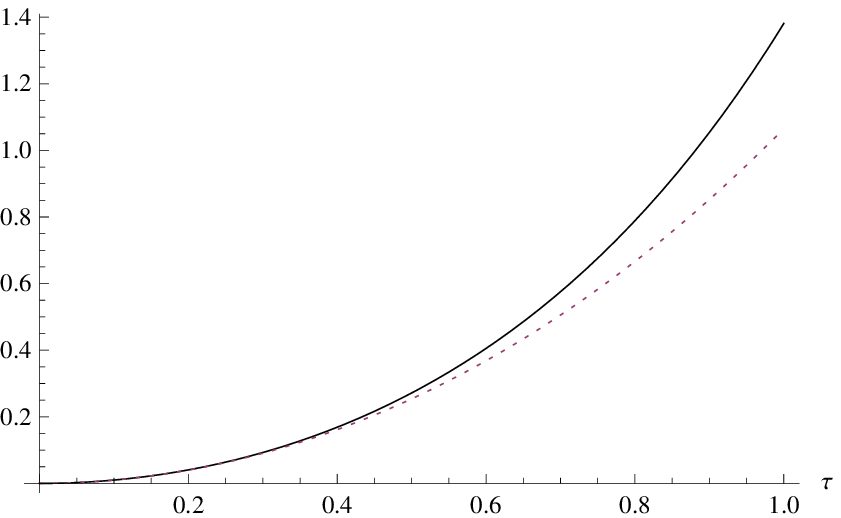}
\includegraphics[height=50mm,keepaspectratio=true,angle=0]{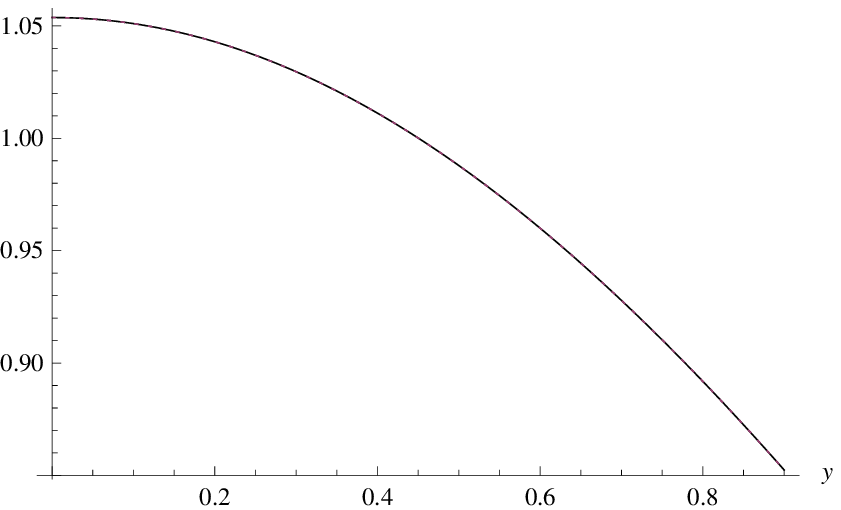}
\caption{\it 
{\bf (a)} Time-dependent average number of photons $\langle {\hat n_j}\rangle$ in mode $j=1,2$ for uncorrelated vacuum inputs: $y=0$ (thick line), $y=0.9$  (dotted line). {\bf (b)}  Average number of photons as function of $y$, for $\tau=0.9$.}
\label{Fig.1}
\end{figure}

For small interaction times $\tau <<1$, $C^2( x \tau) \approx 1$, and the number of photons produced is insensitive to the mismatch $y$. For increasing interaction times the average number of photons in each mode grows exponentially with $C^2( x \tau)$ and $S^2( x \tau)$. For a given interaction time the squeezing parameter $r(y)$ is smaller for bigger mismatch, and we expect smaller production of photons with increasing mismatch. These conclusions are corroborated by the plots of Figure 1.

In Fig.1 (a) is plotted the time-dependent average number of photons in the signal for $ { \bar n}_{10}  = { \bar n}_{20} =  0$:  $y=0$ (thick line) and $y=0.9$ (dotted line). In Fig.1 (b) the the average number of photons is plotted as function of the mismatch parameter $y$, for $\tau=0.9 $. These graphics show that the average number of photons produced by the nonlinear interaction increases with the interaction time $\tau$, and decreases with increasing mismatch $y$.

\section{Degree of entanglement}

The concepts of entanglement and information encoded in a quantum state are closely related. For pure states, bipartite entanglement, is equivalent to the lack of information (mixedness) of the reduced state of each subsystem. For mixed states, each subsystem has its own level of impurity, and moreover the global state is itself characterized by a nonzero mixedness which makes difficult to define a single measure of entanglement.

We assume that our system can be in three different initial states. In the first case, modes $1$ and $2$ are excited from the vacuum state, i.e., the initial state is $|0 \rangle_1 | 0 \rangle_2$, which is the case of the spontaneous parametric downconversion. In the second case the initial state is the product of a coherent state $| \alpha \rangle_1$, of the injected signal, by the vacuum state $| 0 \rangle_2$, for the idler, i.e., the initial state is $| \alpha \rangle_1| 0 \rangle_2$. These two quantum initial states are pure states and the unitary evolution of the system will keep them pure states. 

The third initial state that we consider is a mixed state given by the product of two thermal states in modes 1 and 2, 
\begin{equation}\label{density}
{\hat \rho}_0 = {\hat \rho}_{10} \otimes {\hat \rho}_{20} 
\end{equation}
where 
\begin{equation}\label{densityj}
 {\hat{ \rho}}_{j0}  = \sum_{n_j =0}^{\infty} e^{- \beta_j  n_j } | n_j \rangle \langle n_j |
\end{equation}
is the density operator of mode $j( =1, 2)$ in thermal equilibrium at temperature $T$, with $ \beta_j =\frac {\hbar \omega_j}{k_B T} $ and $k_B$ is the Boltzmann constant. The initial average number of photons in mode $j$ is ${\bar n}_{j0}=\frac{1}{e^{\beta_j}-1}$ . Though at room temperature ${\bar n}_{j0}<<1$ in the optical part of the electromagnetic spectrum, in the microwave part of the spectrum ${\bar n}_{j0}>>1$ and we cannot ignore the presence of thermal photons in the nonlinear medium. 

All these three initial quantum states are Gaussian states, i.e. they have Gaussian Wigner and characteristic functions and hence they are fully characterized by the first and second statistical moments \cite{Simon2000,Werner01} of the dimensionless mode conjugate quadratures $ \hat { x}_j  = \sqrt{\frac{\omega_j}{2 \hbar} }({\hat { a}_j } + {\hat {a}_j^{\dag}  } )$ and $\hat {p}_j =   \frac{1}{{\sqrt  {2 \hbar \omega_j}}}({\hat { a}_j  } - {\hat {a}_j^{\dag}  } )$, $ (j=1,2)$, that is, by the vector of the mean values $ {\vec  X } \equiv  ( \langle {\hat { x}_1 } \rangle , \langle {\hat x}_2 \rangle , \langle {\hat { p}_1 } \rangle,\langle {\hat { p}_2 } \rangle )$ and by the covariance matrix (CM) $  \sigma $ 
\begin{equation}\label{covariance}
\sigma_{ij} \equiv  \frac{1}{2} \langle  {\hat { x}_i  } {\hat { x}_j } + {\hat { x}_j  } {\hat { x}_i } \rangle - \langle  {\hat { x}_i  } \rangle  \langle {\hat { x}_j } \rangle
\end{equation} 
The first moments can be arbitrarily adjusted by local unitary operations. Such operations leave entropy and entanglement invariant. For uncorrelated vacuum inputs, the first moments are already zero, without any need of adjustment. 

The correlations between the quadratures of modes 1 and 2, are encoded in the off diagonal blocks of the CM 
\begin{equation}\label{off}
{\bf \gamma }= \left( 
\begin{array}{cc}
\sigma_{13} & \sigma_{14}  \\ 
\sigma_{23}& \sigma_{24}  \\ 
\end{array}
\right)
\end{equation}

Unitary evolutions, generated by bilinear Hamiltonians such as  (\ref{hamiltonean}), preserve the Gaussian statistics of the quantum states  \cite{Illuminati04,Serafini04}, hence the quantum state of the coupled modes at any instant $t >0$, is still a Gaussian state. 

The degree of entanglement in a pure bipartite system is equivalent to the degree of mixedness of each subsystem and is properly quantified by the entropy of entanglement $\cal E$, defined as the Von Neumann entropy of its reduced density operator  \cite{Simon2000, Werner01}. In terms of the entries of the CM $\sigma$, the Von Neumann entropy of each mode is given by  \cite{Illuminati04}, 
\begin{equation}\label{entropy2}
{\cal E}=  f(\nu_1)
\end{equation}
and
\begin{equation}\label{entropy}
f(x)=(x+ \frac{1}{2}) \ln (x + \frac{1}{2})-(x- \frac{1}{2}) \ln (x- \frac{1}{2})
\end{equation}
where $\nu_j = \sqrt {\sigma_{jj} \sigma_{j+1,j+1}-\sigma_{j,j+1}^{2}} \geq 1/2$  is the symplectic  eigenvalue of the reduced single-mode CM $\sigma_j \,\ (j=1,2)$. Each mode displays maximal disorder subjected to given physical constrains. Our system is unbounded, therefore the entropy is constrained by specifying its mean energy. Whenever $\nu_j=1/2$, mode $j$ is in a pure state and the composite system is in a separable quantum state.

For two uncorrelated vacuum inputs $ | 0 \rangle_1 |0  \rangle_2 $,  the initial average values of photons in each mode is ${\bar n}_{10}= {\bar n}_{20} =0$ and for the initial state $| \alpha \rangle_1 |0 \rangle_2$ they are ${\bar n}_{10}= |\alpha |^2$ and $ {\bar n}_{20} =0$ and the symplectic eigenvalues of the reduced covariance matrices $\sigma_j$ are given by
\begin{equation}\label{symplectic}
\nu_1 = \nu_2 = \frac{1}{2} \sqrt{1 + 4 \frac{ C^2( x \tau) -y^2}{1-y^2}  \frac{ S^2( x \tau) }{1-y^2}(2{ \bar n}_{10} +1)  }
\end{equation}
showing that $\nu_j >1/2 $ for $t>0$, then the entropy of each mode is greater than zero and the two modes are entangled from the beginning of the interaction. For interaction times such that the squeezing parameter is $r(y) << 1$, the entropy of each mode is almost insensitive to the mismatch, for increasing interaction times the exponential growth of the hyperbolic functions dominates and the degree of entanglement decreases with increasing mismatch. This behavior is expressed in Figure 2.

 Figure 2(a)  shows the time-dependent degree of entanglement $\cal E$ for the initial state $ | 0 \rangle_1 |0  \rangle_2 $, for zero mismatch $(y=0)$ (thick line) and for $y=0.9$ (dotted line). In panel (b) the degree of entanglement is plotted as function of the dimensionless mismatch $y$ for a given dimensionless interaction time, $\tau =0.9$. As we expected, the degree of entanglement increases with the interaction time and, for a given interaction time decreases when the mismatch increases. 

\begin{figure}[htb]
\centering
\includegraphics[height=50mm,keepaspectratio=true,angle=0]{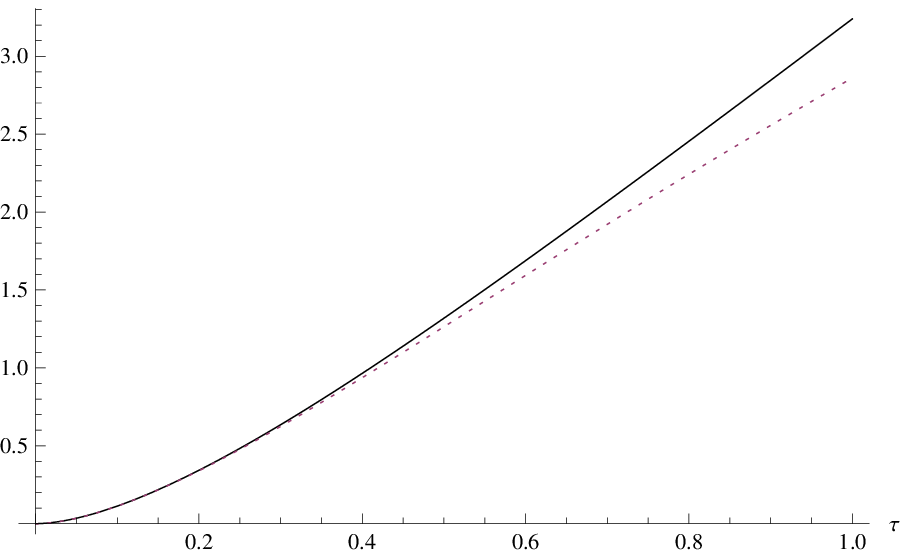}
\includegraphics[height=50mm,keepaspectratio=true,angle=0]{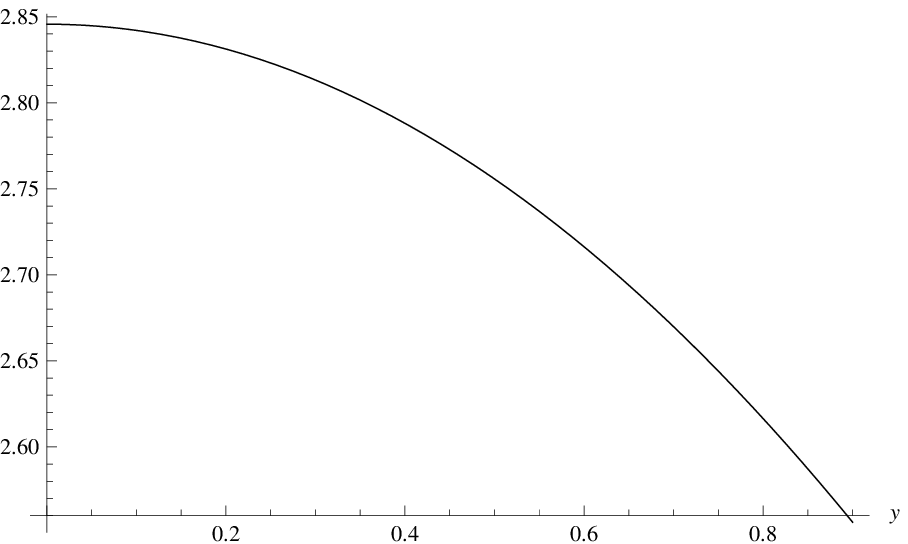}
\caption{\it 
{\bf (a)}Time-dependent degree of entanglement ${\cal E} $ for the initial state $ | 0 \rangle_1 |0  \rangle_2 $ :  $y=0$ (thick line) and $y=0.9$  (dotted line). {\bf (b)} ${\cal E}$ is plotted as function of $y$ for $\tau =0.9$.}
\label{Fig.2}
\end{figure}

In mixed states the mere distinction between classical correlations, i.e. producible by means of local operations and classical communication (LOCC) only, and entanglement, due to a purely quantum interaction between subsystems, is a highly nontrivial task and the entropies of each mode are no more a convenient measure of entanglement.

In our bipartite system, with two interacting modes, we can use the partial transpose (PPT) criterion which says that a quantum state is separable if and only if ${\tilde \nu}_{-} \geq 1/2$, where  ${\tilde \nu}_{-} $ is the smallest symplectic eigenvalue of the partial transpose of $\sigma$, given by \cite{Illuminati04}
\begin{equation}\label{ppt}
{\tilde \nu}_{-} =\sqrt{\frac{1}{2} \left( {\tilde {\Delta}} (\sigma)- \sqrt{{ \tilde {\Delta} }(\sigma)- 4 \det \sigma} \right) }
\end{equation}
where ${\tilde {\Delta}} (\sigma) = \det \sigma_1 + \det \sigma_2 - 2 \det \gamma$. 
The logarithmic negativity provides a proper quantification of entanglement in two-mode Gaussian states and is a computable measure of entanglement, defined by \cite{Horodecki1998, Werner2002}
\begin{eqnarray}
{\cal E}_{\cal N}=   \left\{
 \begin{array}{cc}
 - \ln {2 \tilde \nu}_{-}  & \mbox{if  $\,\,\   - \ln {2 \tilde \nu}_{-}  >0$} \\
 0 & \mbox{if  $\,\,\    - \ln  {2 \tilde \nu}_{-}  <0$ }
 \end{array}
 \right.
 \end{eqnarray}
For the initial uncorrelated thermal state given in (\ref{density}) the entries of the covariant matrix are given in the Appendix and are used to compute the logarithmic negativity ${\cal E}_{\cal N}$ which is plotted in Figure 3.

Figure 3 (a) shows, ${\cal E}_{\cal N}$, as function of time: dashed line corresponds to uncorrelated vacuum input state with $y=0$ and dotted line is the same input state at $y=0.9$, thick line corresponds to the initial thermal state for $y=0$ and the dasheddot line is the same initial condition for $y=0.9$. The function ${\cal E}_{\cal N}$ is always positive for uncorrelated vacuum inputs, independently of the mismatch parameter and for any interaction time, in accordance with our previous conclusions. For the initial thermal state, ${\cal E}_{\cal N}$ becomes positive only after a certain time of interaction. This clearly shows that the entanglement is more difficult to attain for a mixed thermal state. Moreover, the composite system needs more time of interaction to become entangled when the mismatch is nonzero.

In Fig.3 (b) we plot ${\cal E}_{\cal N}$ for the case of the initial thermal state, as function of the mismatch and for $\tau =0.7$. For this interaction time, ${\cal E}_{\cal N}$ is always positive, decreasing with the mismatch meaning that the system is entangled.

Finally, in Fig.3(c) we plot different quantities for the initial thermal state with ${ \bar n}_{10}=1,  { \bar n}_{20}=2$ : the Von Neumann entropy of mode 1 for $y=0$ (dashed line) and for $y=0.9$ (dotted line), and ${\cal E}_{\cal N}$ for $y=0.9$ (thick line). The entropy of mode 1 increases with time and, for zero mismatch, it is always greater than for nonzero mismatch.  

\begin{figure}[htb]
\centering
\includegraphics[height=50mm,keepaspectratio=true,angle=0]{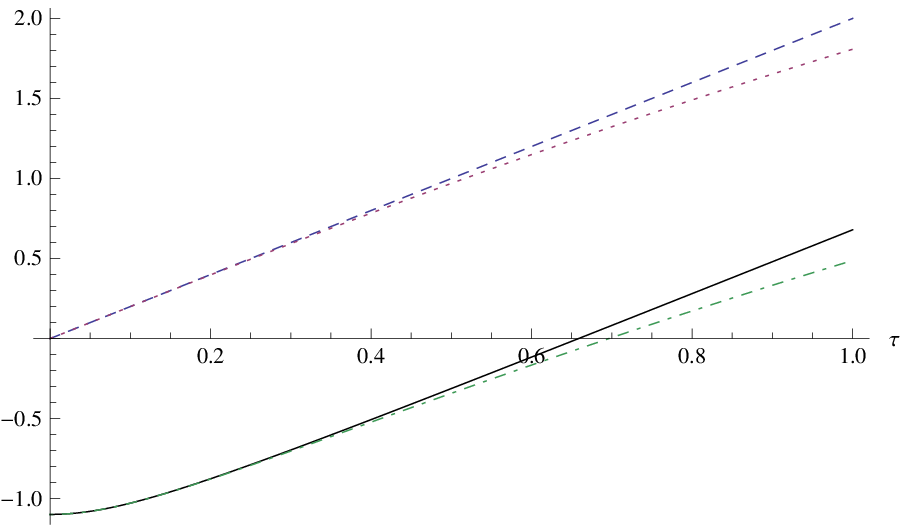}
\includegraphics[height=50mm,keepaspectratio=true,angle=0]{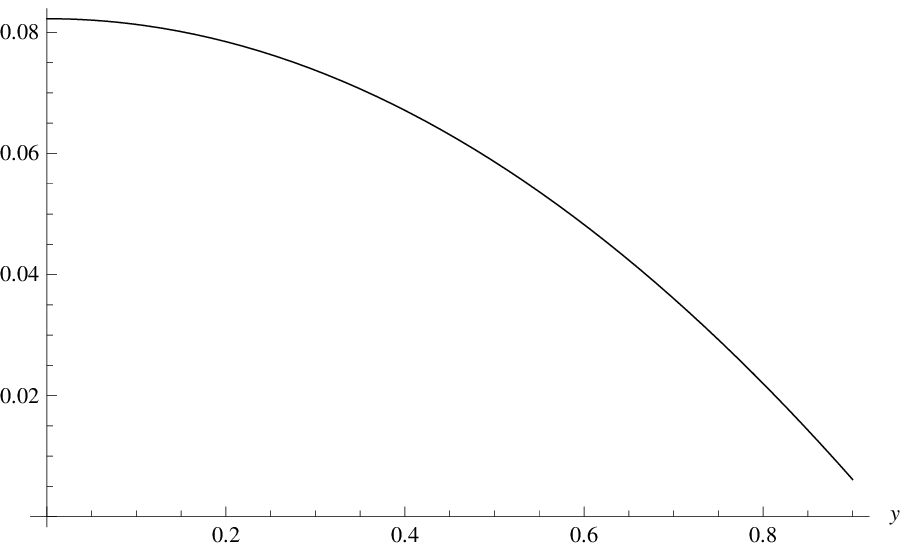}
\caption{\it 
{\bf (a)} ${\cal E}_{\cal N}$ as function of time for the uncorrelated vacuum input state: $y=0$ (dashed line) and  $y=0.9$ (dotted line). The same quantity for the initial thermal: $y=0$ (thick line)  and $y=0.9$ (dot-dashed line) is the same initial condition for $y=0.9$. {\bf (b)} ${\cal E}_{\cal N}$ as function of $y$ for $\tau =0.7$.}
\label{Fig.3}
\end{figure}

\begin{figure}[htb]
\centering
\includegraphics[height=50mm,keepaspectratio=true,angle=0]{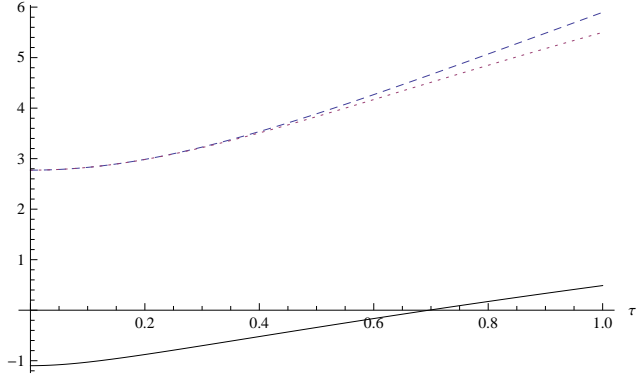}
\caption{\it 
{\bf (c)} Von Neumann entropy of mode 1:  $y=0$ (dashed line) and $y=0.9$ (dotted line), and ${\cal E}_{\cal N}$ for $y=0.9$ (thick  line). All three functions were computed for the initial thermal state (\ref{density}) with ${ \bar n}_{10}=1,  { \bar n}_{20}=2$.}
\label{Fig.3}
\end{figure}

The degree of entanglement for all three initial states that we have studied, decreases with increasing mismatch. This is a direct consequence of the dependence of the effective squeezing parameter $r(y)$, present in the time dependent bosonic operators (\ref{time1}) and (\ref{time2}), with the phase-mismatch.

\section{Summary and concluding remarks}

We have extended the computation of the degree of entanglement, in nondegerate two-mode parametric processes in order to include the effect of a phase mismatch.  A correction factor was introduced in the usual hamiltonean of these processes and exact solutions of the Heisenberg equations were derived. 

It was shown that the new squeezing parameter includes a dependence in the mismatch parameter and that its intensity decreases with increasing mismatch. This behavior is responsible for generating an average number of photons in both modes that decreases with increasing mismatch.

The degree of entanglement was computed for three different initial quantum states. The two pure initial states are associated with the downconversion and with the injected signal of the amplifier. The initial mixed thermal sate takes into account the finite temperature of the nonlinear medium and may be relevant for processes in the microwave region of the electromagnetic spectrum. The degree of entanglement shows a similar behavior in these three cases, i.e, for a given interaction time it decreases with increasing mismatch and increases with the generated number of photon pairs.

To measure the degree of entanglement that we have derived in this paper we propose to make measurements of the statistical properties of the quadratures correlation of the two-modes, which are the entries of the covariance matrix of the system, trough the usual techniques of homodyne detection \cite{Yuen1983, Abbas1983}. 

The Hamiltonean including the phase-mismatch factor is similar to the Hamiltonean associated with a slight detuning between the pump, the signal and the idler, in other words this Hamiltonean takes into account a possible effect of the finite bandwidth of the modes in interaction. In this case, our results suggest a degradation of the quality of entanglement when finite bandwidth of the modes cannot be ignored.  

 \section{Appendix}

Using the time dependent bosonic operators given in eqs.(\ref{time1}) and (\ref{time2}) we compute the entries of the covariance matrix $\sigma$ for the thermal initial state (\ref{density}), obtaining
\begin{equation}\label{sigmajj}
\sigma_{jj} (\tau)   = \langle {\hat n}_j  (\tau)  \rangle  +\frac{1}{2}  \,\,\ ; \,\,\,\  (j=1,2)
\end{equation}
\begin{equation}\label{sigma12}
\sigma_{12} (\tau)   = \sigma_{21} (\tau)  = \sigma_{34} (\tau)  = \sigma_{43} (\tau)  =0  \,\,\ ; \,\,\,\  (j=1,2)
\end{equation}
\begin{equation}\label{sigma13}
\sigma_{24} (\tau)   = \sigma_{13} (\tau)   \,\,\,\ ; \,\,\,\    \sigma_{23} (\tau)   = \sigma_{14} (\tau)
\end{equation}
\begin{equation}\label{sigma13}
\sigma_{13} (\tau)  = ({ \bar n}_{10} + { \bar n}_{20} +1) \left[ \frac{\sin (2  \omega^{\prime}\tau)}{(1-y^2)^{1/2}} C(x \tau)S(x \tau) \\
-  \frac{\cos (2 \omega^{\prime} \tau )y}{1-y^2}S^2 (x \tau)  \right]
\end{equation}
\begin{equation}\label{sigma14}
\sigma_{14} (\tau)   = ({ \bar n}_{10} + { \bar n}_{20} +1) \left[ \frac{\cos (2  \omega^{\prime}x \tau)}{(1-y^2)^{1/2}} C(x \tau)S(x \tau) \\
+  \frac{\sin (2 \omega^{\prime} x \tau )y}{1-y^2}S^2 (x \tau)  \right]
\end{equation}
\\\\\\

\section*{Figures captions}

\noindent {\bf Fig.2} {\bf (a)}Time-dependent degree of entanglement ${\cal E} $ for the initial state $ | 0 \rangle_1 |0  \rangle_2 $ :  $y=0$ (thick line) and $y=0.9$  (dotted line). {\bf (b)} ${\cal E}$ is plotted as function of $y$ for $\tau =0.9$. \newline

\noindent {\bf Fig.3} {\bf (a)} ${\cal E}_{\cal N}$ as function of time for the uncorrelated vacuum input state: $y=0$ (dashed line) and  $y=0.9$ (dotted line). The same quantity for the initial thermal: $y=0$ (thick line)  and $y=0.9$ (dot-dashed line) is the same initial condition for $y=0.9$. {\bf (b)} ${\cal E}_{\cal N}$ as function of $y$ for $\tau =0.7$. {\bf (c)} Von Neumann entropy of mode 1:  $y=0$ (dashed line) and $y=0.9$ (dotted line), and ${\cal E}_{\cal N}$ for $y=0.9$ (thick  line). All three functions were computed for the initial thermal state (\ref{density}) with ${ \bar n}_{10}=1,  { \bar n}_{20}=2$. \newline

\newpage

\end{document}